\documentclass[10pt]{iopart}
\usepackage{amssymb}
\usepackage{epsfig}

\def\beq{\begin{equation}}
\def\eeq{\end{equation}}
\def\beqa{\begin{eqnarray}}
\def\eeqa{\end{eqnarray}}
\def\hf{\textstyle{1\over2}}
\def\3hf{\textstyle{\frac{3}{2}}}

\newcommand{\ket}[1]{\vert #1 \rangle}
\newcommand{\cg}[6]{C^{#5\,#6}_{#1 , #2 ; #3 , #4}}
\newcommand{\sucg}[3]{\left\langle #1 ; #2 \,\vert\, #3 \right\rangle }
\newcommand{\bmat}{\left(\begin{array}}
\newcommand{\emat}{\end{array}\right)}
\newcommand{\bra}[1]{\langle #1 \vert}
\newcommand\unit{\mathinner{\hbox{1}\mkern-4mu\hbox{l}}}
\newcommand\rket[1]{\vert\, #1 )}

\newcommand\ip[2]{\langle #1\,\vert\,#2\rangle}

\renewcommand{\e}{\hbox{\rm e}}
\newcommand{\myfrac}[2]{\leavevmode\kern.1em\raise.5ex\hbox{\scriptsize
$#1$}\kern-.1em {\scriptsize
/}\kern-0.10em\lower.25ex\hbox{\scriptsize $#2$}}
\newcommand{\expect}[1]{\langle\, #1\,\rangle}
\newcommand{\ie}{{\it i.e.\ }}

\begin{document}

\title[su(3) intelligent states]{Su(3) intelligent states as coupled su(3) coherent states}

\author{Benjamin R. Lavoie\footnote{ now at Department of Physics and Astronomy, University of Calgary
Calgary, Alberta, Canada} and Hubert de Guise
\footnote[3]{To whom correspondence should be addressed
(hubert.deguise@lakeheadu.ca)} }
\address{Department of Physics, Lakehead University,
Thunder Bay, ON, P7B 5E1, Canada}

\begin{abstract}
We extend previous work on intelligent states and show how to construct $su(3)$ intelligent states by coupling $su(3)$ coherent states.  We also discuss some properties of the resulting states.
\end{abstract}


\section{Introduction}

The objective of this paper is to show how intelligent states for some observables in the algebra $su(3)$ can be explicitly constructed from coherent states.  Intelligent states
for the observables $\hat \Omega$ and $\hat\Lambda$ are states for which the strict equality 
\begin{equation}
\Delta \Omega\Delta \Lambda =\textstyle\frac{1}{2}\,\vert
\langle[ \hat \Omega,\hat \Lambda ] \rangle \vert ,
\label{robertson}
\end{equation}
holds \cite{aragone}.  A state $\ket{\psi}$ that satisfies Eqn.(\ref{robertson}) also satisfies
\beq
\left(\hat \Omega-i\alpha\hat \Lambda\right)\ket{\psi}=\kappa\ket{\psi}\, ,\qquad
\alpha\in \mathbb{R}\, , \label{eigenvalueproblem}
\eeq
\ie intelligent states are eigenstates of the (non--hermitian) operator $\hat \Omega-i\alpha\hat \Lambda$.

Coherent states are examples of intelligent states for appropriately chosen observables.  It is well known for instance that the 
harmonic oscillator coherent states satisfy 
$\Delta x\Delta p=\hf \hbar$, 
with a similar property also holding for suitably chosen observables evaluated in an angular momentum coherent state.
Coherent states are a special case of intelligent states because, in addition to saturating the uncertainty relation of Eqn.(\ref{robertson}) 
or its angular momentum analogue, the observables also satisfy $\Delta  x=\Delta p$ (in appropriate units).  Intelligent states generalize coherent
states in the sense that Eqn.(\ref{robertson}) still holds but in general $\Delta \Omega\ne \Delta \Lambda$.  

Intelligent states, like coherent states, are not orthogonal.  They were originally introduced for $SU(2)$ by Aragone and collaborators \cite{aragone}.   They were constructed for $SU(2)$ using a non--unitary transformation by Rashid \cite{Rashid}, and using polynomial states in \cite{Milks}.  For $SU(1,1)$ coherent states, the construction often includes solving recursion relations \cite{su11work}.  Properties of $SU(2)$ and $SU(1,1)$ intelligent states have been used in the context of interferometry \cite{intelligentlight}.  

Our work is motivated in part by the resurgence of interest in systems with higher symmetries, such as three-well (or $n$--well) BEC, multiple--path interferometers {\it etc}.  Concurrent with this is the considerable interest by the quantum information community in the study of uncertainty relations and their connection with entanglement  \cite{quantumi} and other non--classical properties of the system.  Intelligent states for observables in higher Lie algebras thus appear to be a "natural" family of states to consider in order to explore such properties.

In addition, the time evolution of angular wave packets associated with diatomic rigid molecules or with a quantum axially-symmetric rigid body has been studied using $su(2)$ intelligent states \cite{Arvieu}.  One could thus envisage that $su(3)$ and $su(n)$ intelligent states could be use to study the time evolution of more general wave packets.

We will show in this paper how $su(3)$ intelligent states can be constructed by coupling together three $SU(3)$ coherent states, 
in direct generalization of the method proposed in \cite{Lavoie} for $SU(2)$ and extended to $SU(1,1)$ in \cite{Joanis}.  We illustrate our method by choosing observables that do not transform by an $su(2)$ subalgebra of $su(3)$, thus bypassing the restrictions found in \cite{daoud}.  Although $SU(3)$ Clebsch-Gordan technology can be quite formidable, the couplings we will require are of the simplest kind as we will restrict our discussion to states in representations of the type $(\lambda,0)$.  The coupling method offers some distinct advantages over other possible constructions as it relies only on known special functions but not on recursion relations \cite{su11work}, and does not hinge on finding a suitable non--linear transformation \cite{Rashid}.   
The coupling method is also immediately generalizable to higher groups once the appropriate coherent states have been found.

Finally, we note that the literature sometimes refer to intelligent states as ``minimum uncertainty states'' \cite{Puri}.  We will stay away from this qualification: strictly speaking, the minimum of $\Delta \Omega$ is $0$ and reached by choosing a normalized eigenstate of $\hat\Omega$, something always possible in a finite--dimensional representation.

\section{Review of $SU(3)$ coherent states}

By restricting our discussion to $su(3)$ irreps of the type $(\lambda,0)$, we may consider the $su(3)$ algebra to be spanned by the eight operators
\beqa
\hat{C}_{ij} &=&\hat a_{i}^{\dag }\hat a_{j},\qquad\qquad  i\neq j=1,2,3, \nonumber\\
\hat h_{1} &=&\hat a_{2}^{\dag }\hat a_{2}-\hat a_{1}^{\dag }\hat a_{1}\, , \\
\hat h_{2} &=&\hat a_{3}^{\dag }\hat a_{3}-\hat a_{2}^{\dag }\hat a_{2}\, ,\nonumber
\eeqa
constructed from harmonic oscillator creation and destruction operators.  
The elements in the algebra act in a natural way on three--dimensional harmonic oscillator states $\ket{n_1n_2n_3}$.  For instance:
\beq
\hat C_{12}\ket{n_1n_2n_3}=\sqrt{(n_1+1)n_2}\,\ket{n_1+1,n_2-1,n_3}\, .
\eeq
A set of basis states for the irrep $(\lambda,0)$ of dimension $\hf(\lambda+1)(\lambda+2)$ is given by
those three--dimensional harmonic oscillator states $\{\ket{n_1n_2n_3}, n_1+n_2+n_3=\lambda\}$.

An element $R(\varpi)$ of the group $SU(3)$, with $\varpi\equiv (\alpha_1,\beta_1,\alpha_2,\beta_2,\alpha_3,\beta_3,\gamma_1,\gamma_2)$
will be written in the form
\beqa
R(\varpi)&=&R_{12}(\alpha_1,\beta_1,-\alpha_1)R_{23}(\alpha_2,\beta_2,-\alpha_2)\,T(\alpha_3,\beta_3,\gamma_1,\gamma_2)\, ,\nonumber \\
T(\alpha_3,\beta_3,\gamma_1,\gamma_2)&=&R_{12}(\alpha_3,\beta_3,-\alpha_3)\e^{i\gamma_1\hat h_1}\e^{i\gamma_2\hat h_2}\, .
\label{su3finite}
\eeqa
Here, $R_{ab}(\vartheta,\varphi,\chi)$ is a $SU(2)$ subgroup transformation mixing modes $a,b$ of the harmonic oscillator. The angles $\beta_i$ range over
$0\le \beta_i \le \pi$. In the fundamental $(1,0)$ representation, the  subgroup transformations have the generic forms
\beq
R_{12}(\omega_{12})=
\left(
\begin{array}{ccc} 
*  & *  & 0 \\
\ast  & *  & 0 \\
0& 0& 1\end{array}\right)\, ,\qquad 
R_{23}(\omega_{23})=
\left(
\begin{array}{ccc} 
1 & 0 &0 \\
0 &* & * \\
0 &* & * \end{array}\right)\, . \label{factors}
\eeq
The $*$ are entries that make the blocks into $SU(2)$ transformations.  The factored form of Eqn.(\ref{su3finite}) can be verified by slightly modifying the factorization algorithm of \cite{su3Dpaper}.

The lowest weight state of the irrep $(\lambda,0)$ is $\ket{00\lambda}$.  It is isotropic under transformations of the type 
$T(\alpha_3,\beta_3,\gamma_1,\gamma_2)$ given in Eqn.(\ref{su3finite}).  Thus we define an $SU(3)$ coherent state
$\ket{\omega}$ as the ``translated'' lowest weight state
\beq
\ket{\omega}\equiv R_{12}(\alpha_1,\beta_1,-\alpha_1)R_{23}(\alpha_2,\beta_2,-\alpha_2)\ket{00\lambda}\, .
\eeq

\section{$SU(3)$ intelligent states}

\subsection{A choice of observables}

Intelligent states are tied to the product of fluctuations of some specified observables.  For $su(3)$, a completely general pair of observables would
be overly complicated, would hide the simplicity of the procedure and would not allow for a comfortable discussion of some of the results.  
With this in mind, we consider the following pair:
\begin{equation}
\hat{{\cal A}}'= \frac{2\pi}{3}\left(\begin{array}{ccc}%
    0 & 0 & 0 \\
    0 & 1 & 0 \\
    0 & 0 & -1
\end{array}\right)
,\quad\hat{{\cal B}}'=\frac{2\pi i}{3\sqrt{3}}\left(\begin{array}{ccc}%
    0 & -1 & 1 \\
    1 & 0 & -1 \\
    -1 & 1 & 0
\end{array}\right),
\end{equation}
with the commutation relation,
\begin{equation}
\hat{{\cal C}}'=-i[\hat{\cal A}',\hat{\cal B}']=\frac{4\pi^2}{9\sqrt{3}}\left(\begin{array}{ccc}%
    0 & 1 & 1 \\
    1 & 0 & -2 \\
    1 & -2 & 0 \\
\end{array}\right).\label{eq:3by3ABcommutetoC}
\end{equation}
Besides being sufficiently simple yet not trivial, the physical motivation behind this choice is connected with 
properties of the eigenstates of ${\cal A}'$ and ${\cal B}'$.
If $\ket{\Psi _{i}^{{\cal A}'}}$ and $\ket{\Phi _{j}^{{\cal B}'}}$
denote any $3$-dimensional eigenvector of $\hat {\cal A}'$ and $\hat {\cal B}',$ respectively, then these
eigenvectors are said to be mutually unbiased \cite{MUB}:%
\begin{equation}
|\ip{\Psi _{i}^{{\cal A}'}}{\Phi _{j}^{{\cal B}'}}|^{2}=\frac{1}{3}.
\end{equation}
In this perspective ${\cal A}'$ and ${\cal B}'$ are direct generalization of the Pauli matrices $\sigma_x$ and $\sigma_y$, the eigenstates of which satisfy the overlap condition $|\ip{\Psi _{i}^{\sigma_x}}{\Phi _{j}^{\sigma_y}}|^{2}=\frac{1}{2}.$

For calculational simplicity, it is convenient to go to a basis where
$\hat{{\cal C}}'=-i[\hat{{\cal A}}',\hat{{\cal B}}']$ is diagonal. This is done through the transformation
\begin{equation}
\hat{U}=\frac{1}{\sqrt{2}}\left(\begin{array}{ccc}%
    0 & \frac{1-\sqrt{3}}{\sqrt{3-\sqrt{3}}} & \frac{1+\sqrt{3}}{\sqrt{3+\sqrt{3}}} \\
    -1 & \frac{1}{\sqrt{3-\sqrt{3}}} & \frac{1}{\sqrt{3+\sqrt{3}}} \\
    1 & \frac{1}{\sqrt{3-\sqrt{3}}} & \frac{1}{\sqrt{3+\sqrt{3}}}
\end{array}\right),\label{eq:3x3diagtrans}
\end{equation}
with the result that our final observables are:
\beqa%
\hat{{\cal A}}=\hat{U}^{-1}\hat{\cal A}'\hat{U}=-\frac{2\pi}{3}\left(\begin{array}{ccc}
    0 & \frac{1}{\sqrt{3-\sqrt{3}}} & \frac{1}{\sqrt{3+\sqrt{3}}} \\
    \frac{1}{\sqrt{3-\sqrt{3}}} & 0 & 0 \\
    \frac{1}{\sqrt{3+\sqrt{3}}} & 0 & 0
\end{array}\right),\nonumber\\
\hat{\cal B}=\hat{U}^{-1}\hat{\cal B}'\hat{U}=\frac{2\pi i}{3}\left(\begin{array}{ccc}
    0 & \frac{1}{\sqrt{3-\sqrt{3}}} & \frac{-1}{\sqrt{3+\sqrt{3}}} \\
    \frac{-1}{\sqrt{3-\sqrt{3}}} & 0 & 0 \\
    \frac{1}{\sqrt{3+\sqrt{3}}} & 0 & 0
\end{array}\right),\\
\hat{\cal C}=\hat{U}^{-1}\hat{\cal C}'\hat{U}=\frac{4\pi^2}{9\sqrt{3}}\left(\begin{array}{ccc}
    2 & 0 & 0 \\
    0 & -1-\sqrt{3} & 0 \\
    0 & 0 & -1+\sqrt{3}
\end{array}\right)\nonumber.
\eeqa%

The equality that defines intelligence, Eqn.(\ref{robertson}), then reads
\beq%
\Delta{\cal A}\Delta{\cal B}=\hf|\langle\hat{\cal C}\rangle|.
\eeq

\subsection{Intelligent states in $(1,0)$.}
 In the three--dimensional space carried by the (fundamental) $(1,0)$ representation, the   eigenvalue problem of Eqn.(\ref{eigenvalueproblem}) for the operator $\hat{\cal
A}-i\alpha\hat{\cal B}$ can be solved to yield the eigenvectors $\ket{\psi^1_k}$ and eigenvalue $\kappa_k$ given by
\beqa%
|\psi^{1}_{1}(\alpha)\rangle &= {\cal N}_1\left(\begin{array}{c}\nonumber
    0 \\
    \frac{1-\sqrt{3}}{\sqrt{2}}\mu^2\\
    1
\end{array}\right)\, ,\quad &\kappa_1=0\, , 
\\
|\psi^{1}_{2}(\alpha)\rangle &= {\cal N}_2\left(\begin{array}{c}
    \sqrt{3+\sqrt{3}}\mu \\
    \sqrt{2+\sqrt{3}}\mu^2 \\
    1
\end{array}\right)\, , \quad &\kappa_2 =-\frac{2\pi}{3}\sqrt{1-\alpha^2}\, , \\
\ket{\psi^{1}_{3}(\alpha)} &= {\cal N}_3\left(\begin{array}{c}\nonumber
    -\sqrt{3+\sqrt{3}}\mu \\
    \sqrt{2+\sqrt{3}}\mu^2 \\
    1
\end{array}\right)\, , \quad &\kappa_3 =\frac{2\pi}{3}\sqrt{1-\alpha^2}\, ,
\eeqa
where 
\beq
\mu=\frac{1+\alpha}{\sqrt{1-\alpha^2}}\in \mathbb{R}\label{su3mudef}
\eeq 
and ${\cal N}_k$ is a normalization constant. 
These eigenvectors can all be identified with coherent states.  Indeed, a generic coherent state in the irrep $(1,0)$ is
of the form
\beqa
\ket{\omega}&=&R_{12}(\alpha_1,\beta_1,-\alpha_1)R_{23}(\alpha_2,\beta_2,-\alpha_2)\ket{001}\nonumber \\
&=&
\left({\renewcommand{\arraystretch}{1.5}\begin{array}{c}
\e^{-i(\alpha_1+\alpha_2)}\sin\hf\beta_1\,\sin\hf\beta_2\\
-\e^{-i\alpha_2}\cos\hf\beta_1\, \sin\hf\beta_2\\
\cos\hf\beta_2
\end{array}}\right)\, .
\eeqa
Thus, we will write
\beq
\ket{\psi^1_k(\alpha)}=R(\omega_k)\ket{001}\, ,\label{intelligentcoherent}
\eeq
where $\omega_k=(\alpha_{1k},\beta_{1k},\alpha_{2k},\beta_{2k})$  are the angles defined so that Eqn.(\ref{intelligentcoherent}) holds.
Simple comparison with the eigenstates $\ket{\psi^1_k}$ produces the correct angles $\alpha_{1k},\beta_{1k},\alpha_{2k},\beta_{2k}$.  
When $\vert\alpha\vert<1$, we immediately see that
$\alpha_{k1}=\alpha_{k2}=0$ always.  The other two angles are
\beq
\begin{array}{ll}
\beta_{11}=0,&\tan\hf\beta_{21}=-\frac{(1-\sqrt{3})\mu^2}{\sqrt{2}}\, ,\\
\tan\hf\beta_{12}=-\frac{\sqrt{3-\sqrt{3}}}{\mu},&\tan\hf\beta_{22}=-\mu\sqrt{3+\sqrt{3}+(2+\sqrt{3})\mu^2}\\
\tan\hf\beta_{13}=\frac{\sqrt{3-\sqrt{3}}}{\mu},&\tan\hf\beta_{23}=-\mu\sqrt{3+\sqrt{3}+(2+\sqrt{3})\mu^2}
\end{array}
\eeq
For $\vert\alpha\vert>1$, the angles are easily determined from the magnitudes and phases of the appropriate ratios of entries.
Thus we have, quite generally,
\beq
\left(\hat{\cal A}-i\alpha\hat{\cal B}\right)\,R(\omega_k)\ket{001}=\kappa_k\,R(\omega_k)\ket{001},
\eeq
with $\kappa_k$ the $k$'th eigenvalue, which is real when $\vert\alpha\vert<1$ but otherwise $0$ or purely imaginary.

\subsection{The coupling property}

Now the eigenvalue problem is \emph{linear} in the generators.  Thus, if $\ket{\psi_i^1(\alpha)}$ and $\ket{\psi_j^1(\alpha)}$ are intelligent, so is the product
$\ket{\psi_i^1(\alpha)}_1\ket{\psi_j^1(\alpha)}_2$ of two ``single particle'' intelligent states.  More formally, if we definite the tensor product states 
\beq
\ket{\psi_i^1(\alpha)}_1\ket{\psi_j^1(\alpha)}_2\equiv \ket{\psi_i^1(\alpha)}_1\otimes \ket{\psi_j^2(\alpha)}_2
\eeq
and the (collective) operators
\beq
{\cal A}={\cal A}_1\otimes \unit_2+\unit_1\otimes {\cal A}_2\, ,\quad 
{\cal B}={\cal B}_1\otimes \unit_2+\unit_1\otimes {\cal B}_2, ,
\eeq
with
\beqa
{\cal A}\ket{\psi_i^1(\alpha)}_1\ket{\psi_j^1(\alpha)}_2&=&\left({\cal A}_1\otimes \unit_2\right)(\ket{\psi_i^1(\alpha)}_1\otimes \ket{\psi_j^1(\alpha)}_2)\nonumber \\
&&+(\unit_1\otimes {\cal A}_2)(\ket{\psi_i^1(\alpha)}_1\otimes \ket{\psi_j^1(\alpha)}_2)\, ,\\
&=&({\cal A}_1\ket{\psi_i^1(\alpha)}_1)\otimes\ket{\psi_j^1(\alpha)}_2+\ket{\psi_i^1(\alpha)}_1\otimes ({\cal A}_2\ket{\psi_j^1(\alpha)}_2)\, , \\
&=&(\kappa_i+\kappa_j)\ket{\psi_i^1(\alpha)}_1\ket{\psi_j^1(\alpha)}_2
\eeqa
we see that the product of intelligent states is also intelligent.

\section{All the intelligent states of $(\lambda,0)$}

We start by noting that the $q$-fold product 
\beq
\ket{00q}=\ket{001}_1\otimes\ket{001}_2\otimes \ldots \otimes \ket{001}_q
\eeq
is the lowest weight for $(q,0)$.  By the previous argument it follows that
\beqa
\rket{\psi^q_k(\alpha)}&=&R(\omega_k)\ket{00q}\\
&=&\left(R(\omega_k)_1\ket{001}_1\right)\otimes\ldots\otimes \left(R(\omega_k)_q\ket{001}_q\right)
\eeqa
is also coherent and simultaneously intelligent (we use the round ket to denote an unnormalized state).

Let us use this to construct the six intelligent states of ${\cal A}$ and ${\cal B}$ for the irrep $(2,0)$.
The eigenvalue problem takes the matrix form
\begin{eqnarray*}%
&&-\frac{2\pi}{3}\left(\hat{\cal A}-i\alpha\hat{\cal B}\right)=\nonumber\\
    &&\bmat{cccccc}
        0&(1-\alpha)\eta_-&0&(1+\alpha)\eta_+&0&0\\
        (1+\alpha)\eta_-&0&(1-\alpha)\eta_-&0&\frac{(1+\alpha)\eta_+}{\sqrt{2}}&0\\
        0&(1+\alpha)\eta_-&0&0&0&0\\
        (1-\alpha)\eta_+&0&0&0&\frac{(1-\alpha)\eta_-}{\sqrt{2}}&(1+\alpha)\eta_+\\
        0&\frac{(1-\alpha)\eta_+}{\sqrt{2}}&0&\frac{(1+\alpha)\eta_-}{\sqrt{2}}&0&0\\
        0&0&0&(1-\alpha)\eta_+&0&0\\
    \emat\,.
\end{eqnarray*}%
with $\eta_+=\sqrt{\frac{2}{3+\sqrt{3}}},$ and $\eta_-=\sqrt{\frac{2}{3-\sqrt{3}}}\,.$ 

Rather than direct diagonalization, we use the coupling property to verify that the product 
\beq
\rket{\psi^2_{k\ell}(\alpha)}=\ket{\psi^1_k(\alpha)}_1\ket{\psi^1_\ell(\alpha)}_2+\ket{\psi^1_\ell(\alpha)}_1\ket{\psi^1_k(\alpha)}_2
\eeq
is contained in the irrep $(2,0)$ and is intelligent with eigenvalue $\kappa_k+\kappa_\ell$.  Clearly there are six such states, symmetric under
permutation of the ``particle index''; by construction they will be eigenstates of 
${\cal A}-i\alpha{\cal B}$ and thus intelligent.  They must therefore be all the intelligent states of $(2,0)$.

More generally, we start with the product
\beq
\rket{\psi_{\lambda_1\lambda_2\lambda_3}(\alpha)}=\left[R(\omega_1)\ket{00\lambda_1}_1\right]\otimes \left[R(\omega_2)\ket{00\lambda_2}_2\right]
\otimes \left[R(\omega_3)\ket{00\lambda_3}_3\right]\, .
\eeq
This state is intelligent by construction but belongs to the decomposable representation $(\lambda_1,0)\otimes(\lambda_2,0)\otimes(\lambda_3,0)$.  We can project it to the irreducible space $(\lambda_1+\lambda_2+\lambda_3,0)$ using Clebsch-Gordan technology.

First, we write
\beq
R(\omega_k)\ket{00\lambda_k}=\sum_{\nu_1\nu_2}\ket{\nu_1\nu_2\nu_3}D^{(\lambda_k,0)}_{\nu_1\nu_2\nu_3;00\lambda_k}(\omega_k)
\eeq
where $\nu_3$ is determined by $\nu_3=\lambda_k-\nu_2-\nu_1$, and $D^{(\lambda_k,0)}_{\nu_1\nu_2\nu_3;00\lambda_k}(\omega_k)$ is the $SU(3)$ $D$-function described in \cite{su3Dpaper} and given by the overlap
\beq
\bra{\nu_1\nu_2\nu_3}R_{12}(\alpha_{1k},\beta_{1k},-\alpha_{1k})R_{23}(\alpha_{2k},\beta_{2k},-\alpha_{2k})\ket{00\lambda_k}\, ,
\eeq
and $\omega_k$ is specified via Eqn.(\ref{intelligentcoherent}).
Next, we couple in sequence:
\beqa
&&\ket{\nu_1\nu_2\nu_3}_1\ket{\mu_1\mu_2\mu_3}_2\ket{\tau_1\tau_2\tau_3}_3\nonumber \\
&&=\ket{\nu_1\nu_2\nu_3}_1\ket{M_1M_2M_3}_{23}\
\sucg{\mu_1\mu_2\mu_3}{\tau_1\tau_2\tau_3}{M_1M_2M_3}
\, , \\
&&=\ket{N_1N_2N_3}
\sucg{\nu_1\nu_2\nu_3}{M_1M_2M_3}{N_1N_2N_3}
\sucg{\mu_1\mu_2\mu_3}{\tau_1\tau_2\tau_3}{M_1M_2M_3}\
\eeqa
where $\sucg{a_1a_2a_3}{b_1b_2b_3}{c_1c_2c_3}$ is the $SU(3)$ CG coefficient  $\cg{(q,0)}{a_1a_2a_3}{(p,0)}{b_1b_2b_3}{(p+q,0)}{c_1c_2c_3}$.  Note that because
all kets (including those occurring in the intermediate and final coupling) belong to an irrep of the type $(q,0)$ that does not have any weight multiplicity,
a simple listing of the triple $n_1n_2n_3$ is enough to uniquely identify the state and its weight.  The product does not contain any sum because
the weights of the intermediate and final states are completely specified by the weights of the initial states.  Hence, we denote 
by $\ket{\psi^{\lambda_1+\lambda_2+\lambda_3}_{\lambda_1\lambda_2\lambda_3}}$
the intelligent state 
of irrep $(\lambda_1+\lambda_2+\lambda_3,0)\equiv (\lambda,0)$ 
constructed from the coupling of $\ket{\psi_1^{\lambda_1}}\ket{\psi_2^{\lambda_2}}\ket{\psi_3^{\lambda_3}}$.  Its explicit 
expression is given by
\beqa
&&\rket{\psi^{\lambda}_{\lambda_1\lambda_2\lambda_3}(\alpha)}=\sum_{N_1N_2}\ket{N_1N_2N_3}F(N_1N_2N_3)\nonumber \\
&&F(N_1N_2N_3) =\sum_{\nu_2}\sucg{0\nu_2\nu_3}{M_1M_2M_3}{N_1N_2N_3}D^{(\lambda_1,0)}_{0\nu_2\nu_3;00\lambda_1}(\omega_1)
\nonumber \\
&&\times  \sum_{\mu_1\mu_2}
\sucg{\mu_1\mu_2\mu_3}{\tau_1\tau_2\tau_3}{M_1M_2M_3}\, D^{(\lambda_2,0)}_{\mu_1\mu_2\mu_3;00\lambda_2}(\omega_2)
D^{(\lambda_3,0)}_{\tau_1\tau_2\tau_3;00\lambda_3}(\omega_3) \, , 
\label{intelligentfinal}
\eeqa
with $\lambda=\lambda_1+\lambda_2+\lambda_3$, $\tau_k=M_k-\mu_k=N_k-\mu_k-\nu_k$.

The $SU(3)$ $D$--functions are given in terms of $SU(2)$ $d$-functions \cite{su3Dpaper}:
\beqa
D^{(\lambda_1,0)}_{\nu_1\nu_2\nu_3;00\lambda_1}(\omega_1)&=&\delta_{\nu_1 0}\
d^{\frac{1}{2}\lambda_1}_{\frac{1}{2}(\nu_1+\nu_2-\nu_3),-\frac{1}{2}\lambda_1}(\beta_{21}) \nonumber \\
D^{(\lambda_2,0)}_{\mu_1\mu_2\mu_3;00\lambda_2}(\omega_2)&=&
d^{\frac{1}{2}(\mu_1+\mu_2)}_{\frac{1}{2}(\mu_1-\mu_2),-\frac{1}{2}(\mu_1+\mu_2)}(\beta_{12})
d^{\frac{1}{2}\lambda_2}_{\frac{1}{2}(\mu_1+\mu_2-\mu_3),-\frac{1}{2}\lambda_2}(\beta_{22})\, ,\nonumber \\
D^{(\lambda_3,0)}_{\tau_1\tau_2\tau_3;00\lambda_3}(\omega_3)&=&
d^{\frac{1}{2}(\tau_1+\tau_2)}_{\frac{1}{2}(\tau_1-\tau_2),-\frac{1}{2}(\tau_1+\tau_2)}(\beta_{13})
d^{\frac{1}{2}\lambda_3}_{\frac{1}{2}(\tau_1+\tau_2-\tau_3),-\frac{1}{2}\lambda_3}(\beta_{23})\, ,\nonumber \\
\eeqa
where
\beq
d^J_{M,-J}(\beta)=\sqrt{\frac{(2J)!}{(J+M)!(J-M)!}}\left(\cos\hf\beta\right)^{J-M}\,\left(-\sin\hf\beta\right)^{J+M}\, .
\eeq
The $SU(3)$ Clebsch-Gordan coefficient $\sucg{n_1n_2n_3}{m_1m_2m_3}{n_1+m_1,n_2+m_2,n_3+m_3}$ is easily evaluated as
\beqa
&&\sucg{n_1n_2n_3}{m_1m_2m_3}{n_1+m_1,n_2+m_2,n_3+m_3}\nonumber \\
&&\qquad = \sqrt{\frac{p!q!}{(p+q)!}\frac{(n_1+m_1)!}{n_1!m_1!}\frac{(n_2+m_2)!}{n_2!m_2!}\frac{(n_3+m_3)!}{n_3!m_3!}}
\eeqa
subject to the constraints $n_1+n_2+n_3=p$, $m_1+m_2+m_3=q$.

By construction, $\rket{\psi^{\lambda}_{\lambda_1\lambda_2\lambda_3}(\alpha)}$ is intelligent and belongs to the irrep $(\lambda,0)$ 
(albeit not correctly normalized).  By simply going over all those (positive, integer) values of $\lambda_k$ such that $\lambda=\lambda_1+\lambda_2+\lambda_3$, we find $\hf(\lambda+1)(\lambda+2)$ different linearly independent intelligent states.   As the dimension of $(\lambda,0)$ is precisely $\hf(\lambda+1)(\lambda+2)$, it must be that all the intelligent states of this representation are of the form given in Eqn.(\ref{intelligentfinal}).  If all but one of the $\lambda_i$ are $0$, then the state is an $SU(3)$ coherent state.

\section{Selected Results}%

The $su(3)$ intelligent states are the solutions to the eigenvalue equation
\beq%
(\hat{\cal A}-i\alpha\hat{\cal B})\ket{\psi^{\lambda}_{\lambda_1\lambda_2\lambda_3}(\alpha)}=\lambda\ket{\psi^{\lambda}_{\lambda_1\lambda_2\lambda_3}(\alpha)}\,,
\eeq%
and they have the eigenvalues
\beq%
\lambda=\frac{2\pi}{3}\sqrt{1-\alpha^2}\ (\lambda_3-\lambda_2)\,.
\eeq%
One also shows that
\beq%
(\Delta{\cal A})^2=-\frac{1}{2}\alpha\expect{\hat{\cal C}}, 
\qquad(\Delta{\cal B})^2=-\frac{1}{2\alpha}\expect{\hat{\cal C}}\,.
\eeq%

The uncertainty curves for the $su(3)$ intelligent states display an expected behaviour: for $\alpha=0$, the uncertainty is zero as the states are eigenstates of 
${\cal A}$.  For $\alpha=\pm\infty$, they are eigenstates of ${\cal B}$ so the uncertainty goes to zero again.  There are
discontinuities at $\alpha=\pm1$.  Despite extensive numerical experiments, we have not been able to
determine a trend where the uncertainty, overall, is higher or
lower for states of a given $\lambda$ with different values of $\lambda_1,\,\lambda_2,\,\lambda_3$.

\begin{figure}[htb]
\includegraphics[scale=0.65]{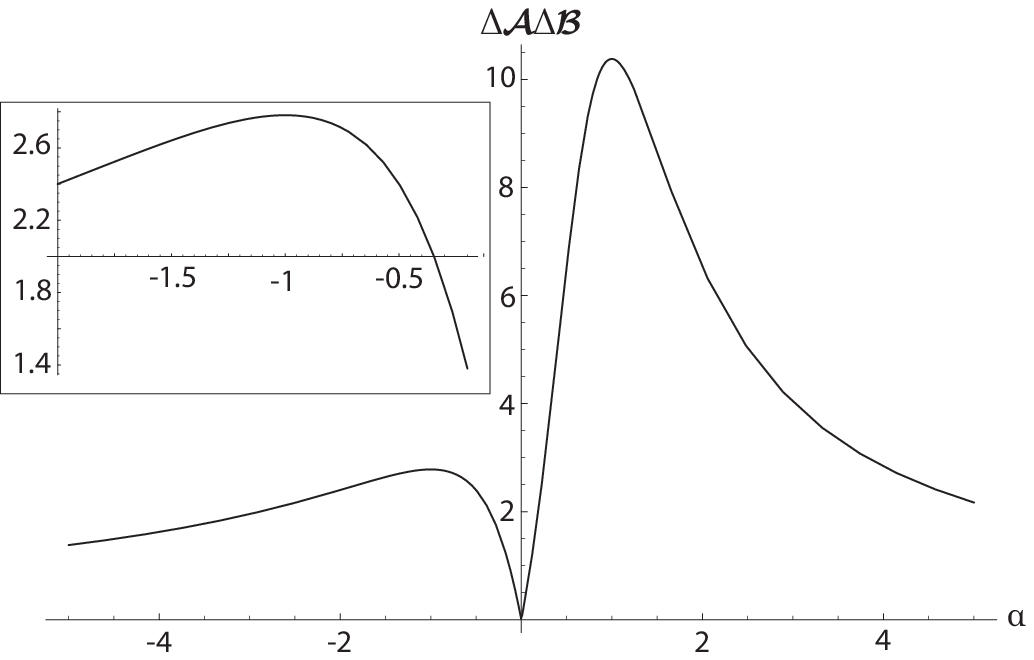}\quad \includegraphics[scale=0.65]{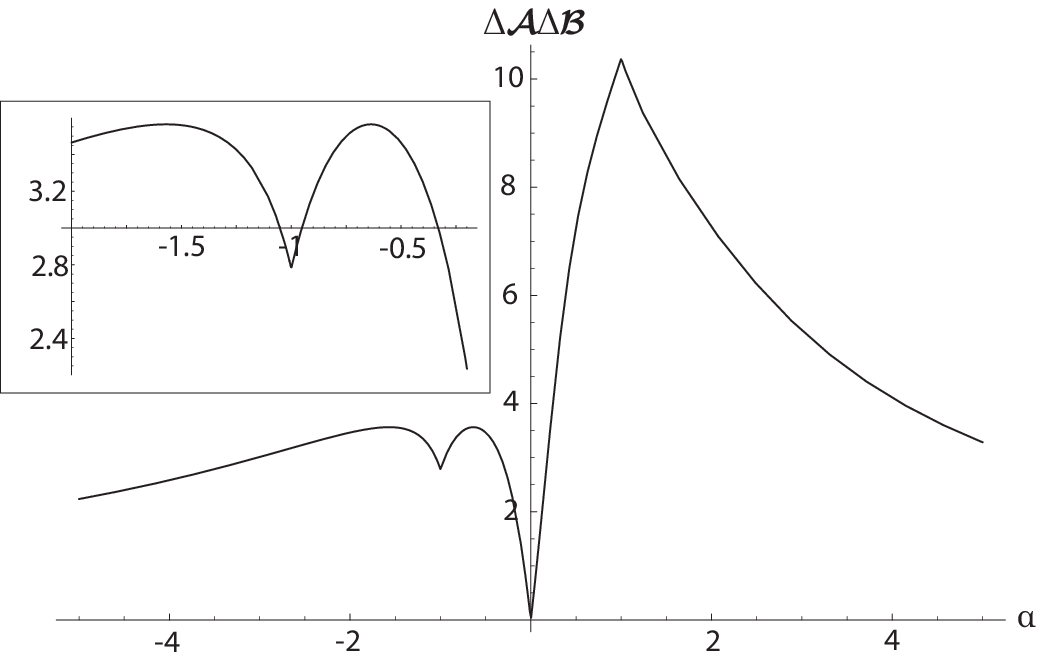}
    \caption{Two plots of $\Delta{\cal A}\Delta{\cal B}$ for $\lambda=3$. The inset is an expanded
    view around $\alpha=-1$. \textbf{Left:} $\lambda_1=3,\,\lambda_2=0,\,\lambda_3=0$, \textbf{Right:} $\lambda_1=1,\,\lambda_2=2,\,\lambda_3=0$.} \label{fig:su3lambeq3}
\end{figure}

\begin{figure}[htb]
\includegraphics[scale=0.5]{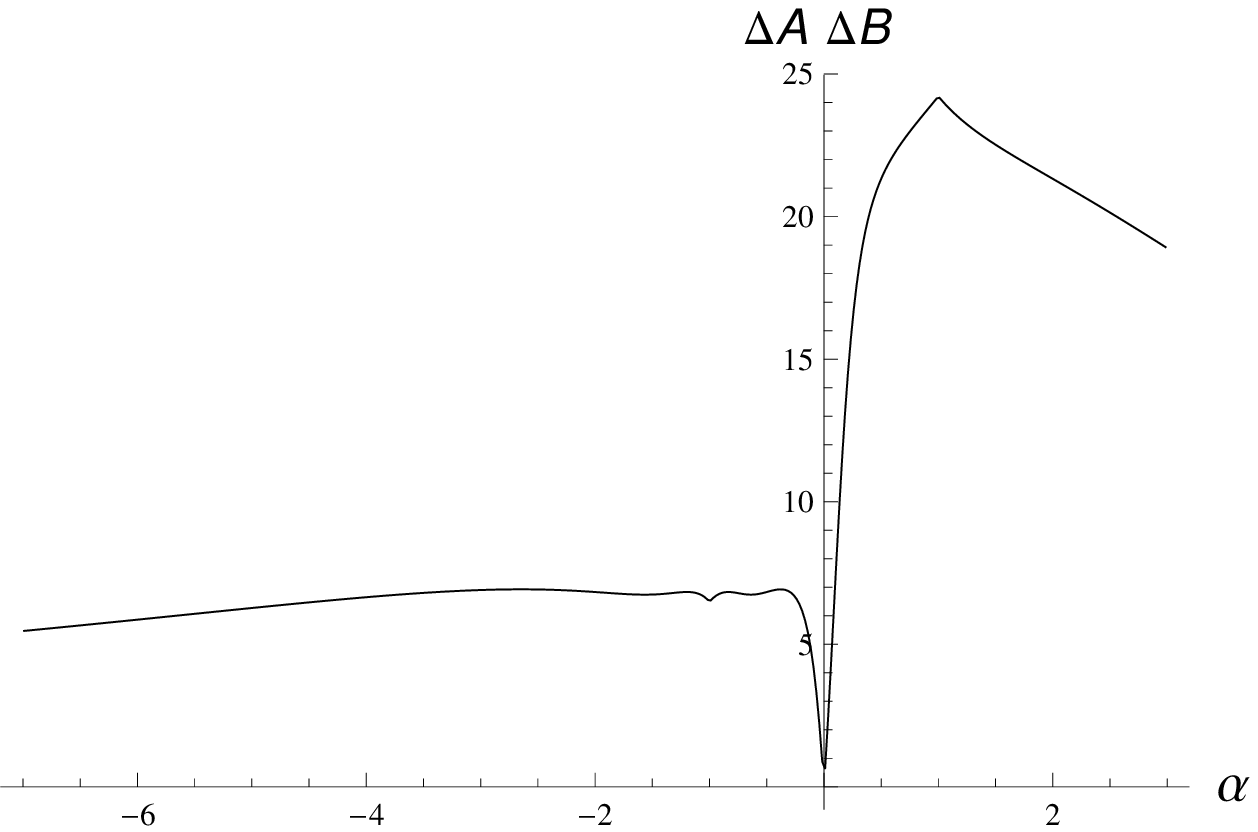}\qquad\includegraphics[scale=0.5]{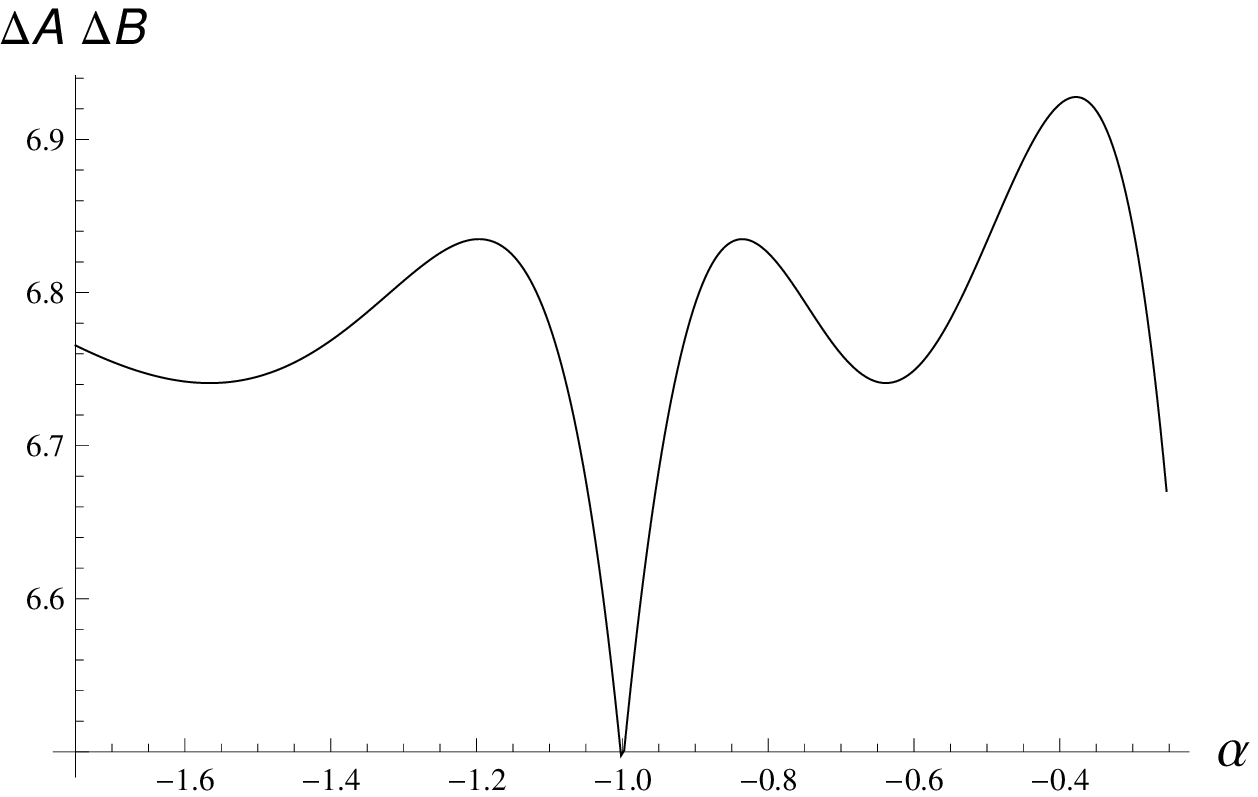}
    \caption{Two plots of $\Delta{\cal A}\Delta{\cal B}$ for $\lambda_1=2,\,\lambda_2=4,\,\lambda_3=1$. This state is one of 36 intelligent state in the irrep $(7,0)$. The rightmost figure gives details of the curve near $\alpha=-1$.} \label{fig:su3lambeq7}
\end{figure}

Figures~\ref{fig:su3lambeq3} and~\ref{fig:su3lambeq7} illustrate typical uncertainty curves for the $su(3)$
states.

A striking feature of the su(3) graphs is the difference in amplitude between positive and negative
$\alpha$. The overall uncertainty for negative $\alpha$ is significantly less than that for positive $\alpha$
for every graph produced up to this point. However, the height of the graph at $\alpha=-1$ can easily be
determined. From the definition of $\mu$, Eq.(\ref{su3mudef}), it can be shown that
\beq%
\lim_{\alpha\rightarrow-1}\frac{1+\alpha}{\sqrt{1-\alpha^2}}=0.
\eeq%
From this, we deduce that the angles in the transformations $\omega_k$ go to $0$ as $\alpha\to -1$.   The uncertainty is then simply
\beq
\Delta{\cal A}\Delta{\cal B} = \hf|\langle\hat C\rangle|
        =\displaystyle\frac{2\pi^2(\sqrt{3}-1)\lambda}{9\sqrt{3}}.
\eeq
Thus, for any $\lambda$ the uncertainty is easily determined for $\alpha=-1$.
Due to the relative complexity of the general expression for the $su(3)$ intelligent states, it was
not possible to establish other analytical results.  It should be noted, however, that the curves presented above do not change if 
$\lambda_2$ and $\lambda_3$ are interchanged.

\section{Conclusion}

This paper shows how one can construct intelligent states for $su(3)$ observables using as ingredients $SU(3)$ coherent states.  It is clear that the method, originally
developed for $su(2)$ and $su(1,1)$ and here applied to $su(3)$, can be generalized to $su(n)$ intelligent states.  For the simplest symmetric (or one--rowed) representations, the coupling coefficients required are easy to calculate so the construction is immediate, and not limited as to the choice of observables.  The CG coefficients and the simple form of the group functions given in Eqns. (\ref{su3finite}) and (\ref{factors}) guarantee that the coherent states are properly normalized.  One must, in the end, properly normalize the intelligent state but this (numerical) procedure and the extraction of relevant matrix elements remains simpler than the expressions obtained using the polynomial method of \cite{Milks}.  It is also applicable to the construction of intelligent states for any representations, something that becomes complicated using polynomial states.  

This work was supported by NSERC of Canada.  The work of BRL was supported by the Government of Ontario through its OGS award program.

\end{document}